\documentclass[numbers,webpdf,imaiai]{ima-authoring-template}%

\usepackage{booktabs}
\usepackage{float}
\usepackage{multirow}
\graphicspath{{Fig/}}
\usepackage{soul}
\usepackage{amsmath}
\usepackage{array}

\theoremstyle{thmstyletwo}%
\numberwithin{equation}{section}

\begin{document}

\copyrightyear{2025}
\vol{00}
\pubyear{2025}
\access{Advance Access Publication Date: Day Month Year}
\appnotes{Paper}
\copyrightstatement{Preprint submitted to SOCO 2024-IGPL Special Issue from the International Journal of the IGPL}
\firstpage{1}


\title[Optimization of embeddings storage for RAG systems.]{\textbf{Optimization of embeddings storage for RAG systems using quantization and dimensionality reduction techniques.}}

\author{Naamán Huerga-Pérez, Rubén Álvarez, Rubén Ferrero-Guillén, Alberto Martínez-Gutiérrez, and Javier Díez-González
\address{\orgdiv{Department of Mechanical, Computer and Aerospace Engineering}, \orgname{Universidad de León}
\orgaddress{\postcode{24071}, \state{León}, \country{Spain}}}}

\authormark{Naamán Huerga-Pérez et al.}

\corresp[*]{Corresponding author: \href{email:javier.diez@unileon.es>}{javier.diez@unileon.es}}

\received{30}{4}{2025}
\revised{Date}{0}{Year}
\accepted{Date}{0}{Year}


\abstract{Retrieval-Augmented Generation enhances language models by retrieving relevant information from external knowledge bases, relying on high-dimensional vector embeddings typically stored in \textit{float32} precision. However, storing these embeddings at scale presents significant memory challenges. To address this issue, we systematically investigate on MTEB benchmark two complementary optimization strategies: quantization, evaluating standard formats (\textit{float16}, \textit{int8}, \textit{binary}) and low-bit floating-point types (\textit{float8}), and dimensionality reduction, assessing methods like PCA, Kernel PCA, UMAP, Random Projections and Autoencoders. 
Our results show that \textit{float8} quantization achieves a 4x storage reduction with minimal performance degradation (\(<0.3\%\)), significantly outperforming \textit{int8} quantization at the same compression level, being simpler to implement. PCA emerges as the most effective dimensionality reduction technique. Crucially, combining moderate PCA (e.g., retaining 50\% dimensions) with \textit{float8} quantization offers an excellent trade-off, achieving 8x total compression with less performance impact than using \textit{int8} alone (which provides only 4x compression). 
To facilitate practical application, we propose a methodology based on visualizing the performance-storage trade-off space 
to identify the optimal configuration that maximizes performance within their specific memory constraints. 
}




\maketitle

\section{Introduction}

Large Language Models (LLMs) have demonstrated remarkable capabilities across a wide spectrum of Natural Language Processing (NLP) tasks, promoting changes in our interactions with information and technology. However, their effectiveness can be constrained by inherent limitations. LLMs possess static knowledge, frozen during their last training, rendering them unaware of recent events or newly emerged information \cite{ragsurvey}. Furthermore, they often lack access to private or domain-specific data crucial for relevant real-world applications, particularly in enterprise settings. Additionally, LLMs can exhibit \textit{hallucinations}, generating plausible-sounding but factually incorrect or nonsensical statements that compromise their reliability \cite{llm_hallucinations}.

To overcome these limitations, Retrieval-Augmented Generation (RAG) has emerged as a highly effective framework \cite{ragfacebook}. RAG enhances LLMs by dynamically retrieving relevant information from external knowledge sources before the text generation process begins. This retrieved information, typically in the form of text snippets or documents, is then provided as additional context to the LLM along with the original user prompt. This grounding mechanism allows RAG systems to synthesize responses that are not only contextually appropriate but also informed by timely and specific external knowledge \cite{ragsurvey}.

The advantages offered by RAG are significant and drive its adoption across diverse fields. Firstly, it enables LLMs to access and incorporate up-to-date information, avoiding the need for frequent and costly model retraining \cite{ragsurvey}. Secondly, RAG facilitates the adaptation of general-purpose LLMs to specific domains (e.g., legal, medical, financial) or proprietary enterprise knowledge bases, enhancing their relevance and utility without compromising data privacy during training \cite{ft_vs_rag}. Thirdly, by grounding responses in retrieved evidence, RAG significantly reduces the propensity for \textit{hallucinations}, leading to more factual and trustworthy outputs \cite{ragfacebook}. This grounding also provides provenance for the generated information, as the source documents can often be cited or inspected.

Consequently, RAG is being successfully applied to a wide range of knowledge-intensive NLP tasks. It has set state-of-the-art results in open-domain question answering, where systems must find answers within vast corpora \cite{ragfacebook}. RAG architectures are improving dialogue systems by providing agents with access to relevant conversational history or external facts \cite{ragsurvey}. It is also used for abstractive, fact verification, and even in scenarios involving less popular or low-frequency knowledge where standard LLMs typically struggle \cite{ft_vs_rag}. 

Central to the RAG architecture is the concept of embeddings and the use of specialized vector databases \cite{ragsurvey}. 
These embeddings are stored in full \textit{float32} precision and indexed for efficient similarity search. However, the high dimensionality of embeddings generated by state-of-the-art models (often 1536 dimensions or more) presents a significant practical challenge: storing these vectors at scale demands substantial memory resources. 

While compression techniques like quantization and dimensionality reduction exist, their application and interaction within RAG systems, especially concerning novel low-bit formats (e.g. \textit{float8}) and diverse reduction methods, requires further investigation. Thus, this paper addresses this gap by systematically analyzing the impact of various quantization strategies and multiple dimensionality reduction techniques, both individually and combined. We quantify the trade-offs between storage compression and retrieval performance to establish effective strategies and propose a methodology for selecting optimal configurations based on specific memory and performance constraints.

Therefore, the main contributions of this work are:
\begin{itemize}[leftmargin=*]
    \item An evaluation of the impact on the RAG retrieval performance of various quantization data types, including standard formats (\textit{float16}, \textit{int8}, \textit{binary}) and novel low-bit floating-point formats, such as \textit{float8} variants commonly used in deep learning model optimization.
    \item An investigation on the effectiveness of applying a diverse set of dimensionality reduction techniques (PCA, Kernel PCA, UMAP, Autoencoders, Random Projections) to embedding vectors, analyzing their impact on storage requirements and retrieval quality.
    \item A systematic study of the combined effects and interactions when applying both quantization (using different data types) and various dimensionality reduction techniques simultaneously to RAG embeddings.
    \item The proposal of a methodology to select the optimal combination of a dimensionality reduction technique and a quantization format maximizing retrieval performance 
    under specific, predefined memory constraints for embeddings storage.
\end{itemize}

\section{Related Works}\label{sec2}

The challenge of efficiently storing and retrieving the high-dimensional embedding vectors inherent to RAG systems has spurred significant research into various compression techniques. These approaches predominantly fall into two main categories: quantization, which reduces the number of bits used to represent each component of a vector, and dimensionality reduction, which decreases the total number of components per vector, maximizing the information retained in the remaining dimensions. Embeddings are typically generated and stored using standard 32-bit floating-point numbers (\textit{float32}) \cite{ieee754}, establishing a baseline against which the efficacy and trade-offs of compression techniques are measured.

Quantization is arguably the most widely adopted strategy, drawing heavily from methods established in the broader field of deep learning model compression \cite{model_compression_overview}. Standard formats like \textit{float16} provide a straightforward 2x storage reduction compared to \textit{float32}. Scalar quantization, often involving the conversion of \textit{float32} values to 8-bit integers (\textit{int8}), achieves a 4x compression factor. This method can benefit from optimized CPU instructions like SIMD for accelerated similarity comparisons, although it typically necessitates a data-dependent calibration step to map the floating-point range to the integer range, potentially incurring some loss in retrieval accuracy \cite{Shakir2024}. More aggressive techniques include binary quantization, where each vector component is reduced to a single bit (\textit{0} or \textit{1}), yielding a 32x compression factor and enabling extremely fast comparisons using bitwise operations. However, the substantial precision loss associated with binary quantization often requires a subsequent re-scoring step. As demonstrated by Shakir et al. (2024) \cite{Shakir2024}, re-ranking the initial candidates retrieved using binary embeddings with higher-precision vectors is crucial to maintain acceptable performance, allowing recovery from approximately 92.5\% to around 96\% of the baseline performance in their specific experiments. Despite these advancements, much of the existing common practices focus on the established \textit{int8} and \textit{binary} formats, leaving the potential of newer low-bit floating-point formats largely unexplored in the context of embedding retrieval.

Dimensionality reduction presents an alternative or complementary path to compression. Techniques like PCA are applied post-hoc to reduce the number of dimensions stored, aiming to preserve the most significant variance in the data. Wang (2019) \cite{embeddings_apple_pca} specifically proposed leveraging PCA not merely for reduction but as an efficient method to select the optimal embedding dimensionality. Their approach involves training a single high-dimensional embedding model initially, then using PCA to evaluate the performance of subsets corresponding to principal components, thereby avoiding the need to train multiple models of varying dimensions. Other techniques, such as Matryoshka Representation Learning (MRL) proposed by Kusupati et al. (2022) \cite{kusupati2024matryoshkarepresentationlearning}, integrate the concept of variable dimensionality directly into the training phase. MRL optimizes nested, lower-dimensional representations within the high-dimensional embedding vector during training, allowing for the selection of an appropriate embedding size post-training. This contrasts with the purely post-hoc nature of standard PCA application but requires specific modifications to the model training pipeline. A simpler strategy involves training models designed inherently to produce lower-dimensional outputs through evaluations, such as those facilitated by benchmarks like MTEB \cite{mteb_arxiv_paper}, which proves this methodology might compromise some representational capacity compared to higher-dimensional counterparts. Regardless of the method, reducing dimensions inherently risks discarding information that might be crucial for capturing fine-grained semantic distinctions necessary for specific retrieval tasks.

While these quantization and dimensionality reduction techniques offer valuable tools for optimizing embedding storage, existing research often evaluates them in isolation. However, understanding their interaction is crucial, as they address the fundamental trade-off inherent in embedding compression: maintaining high retrieval performance, which typically benefits from higher precision and dimensionality, versus minimizing storage costs, which can necessitate lower precision and fewer dimensions. For this purpose, the combined application of these methods offers a wide spectrum of possibilities to balance these competing factors. 

Additionally, the specific interplay, particularly when combining novel quantization formats like \textit{float8} with diverse dimensionality reduction techniques (such as PCA, Kernel PCA, UMAP, Autoencoders, and Random Projections), has not yet been analyzed in the literature of RAG embedding retrieval. Therefore, this work aims to bridge this gap by providing a comprehensive analysis of both individual and combined compression effects. Furthermore, we propose a structured methodology to navigate the resulting complex performance-storage trade-offs, facilitating the selection of optimal compression strategies tailored to specific performance requirements and memory constraints.

\vspace{-3mm}
\section{Problem Definition}\label{sec3}

RAG systems enhance the capabilities of LLMs by dynamically incorporating relevant, often external and up-to-date, information into the generation process \cite{ragfacebook}. 

The fundamental idea is that texts with similar meanings are mapped to vectors that are close to each other in the embedding space \cite{reimers2019sentenceembeddings}. These embeddings, along with their corresponding text, are stored and indexed in specialized Vector Databases optimized for efficient retrieval based on vector similarity \cite{ragsurvey}. When a user query is received, its embedding is computed, and the database is searched to find the $k$ most similar document embeddings. The associated text snippets are then used to augment the input prompt for a LLM, providing relevant context for generating an informed response \cite{ragfacebook}.

However, the practical implementation and scalability of RAG systems face a significant challenge related to the storage of these embedding vectors. State-of-the-art embedding models frequently generate vectors with very high dimensionality – common methods include 1536 or 3072 dimensions, while other leading models evaluated on benchmarks such as MTEB can reach  4096 dimensions or even higher values \cite{mteb_arxiv_paper}. While there is a general trend suggesting that higher dimensionality often correlates with better representational capacity and thus improved retrieval performance on benchmarks \cite{mteb_arxiv_paper}, this directly translates into substantial storage requirements.

These vectors are typically stored using standard 32-bit single-precision floating-point numbers (\textit{float32}), where each dimension consumes 4 bytes of memory \cite{ieee754}. Consequently, storing large collections of high-dimensional embeddings demands significant memory resources for efficient retrieval (loading embeddings from disk introduces significant latency). For instance, a knowledge base containing just one million documents embedded into 1536-dimensional \textit{float32} vectors would require approximately 
6.1 \text{ GB} of RAM, excluding any overhead 
or indexing structures. As datasets scale to tens or hundreds of millions of documents, the memory footprint can 
escalate to hundreds of gigabytes.

This substantial memory requirement leads to significant operational costs, especially in cloud environments where RAM is often a primary cost factor. Furthermore, it can render the deployment of RAG systems infeasible on resource-constrained platforms, such as edge devices, mobile applications, or even standard browser runtimes \cite{Shakir2024}.

Mitigating these storage costs without excessively compromising the quality of retrieval performance is therefore crucial for the widespread adoption and efficient operation of RAG systems. This motivates the exploration of compression techniques specifically tailored for embedding vectors. This paper investigates two complementary optimization strategies (i.e., quantization and dimensionality reduction).

\begin{figure}[h]
    \centering
    \includegraphics[width=0.9\textwidth]{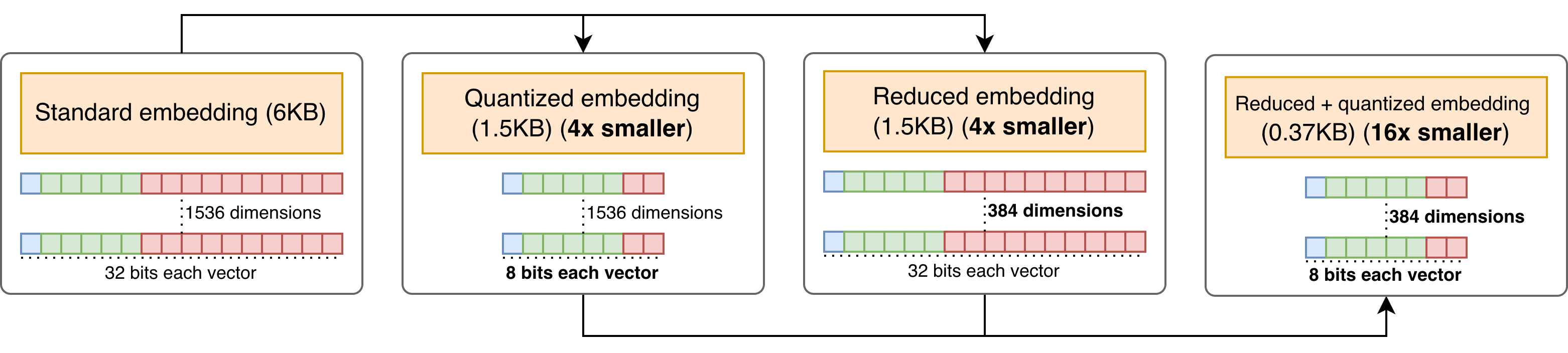}
    \caption{Illustration of embedding storage optimization techniques. From left to right: (1) Standard embedding using 1536 dimensions and 32 bits per dimension (6 KB). (2) Quantized embedding, reducing precision to 8 bits per dimension (1.5 KB, 4x smaller). (3) Reduced embedding, decreasing dimensionality to 384 dimensions (1.5 KB, 4x smaller). (4) Combined reduced and quantized embedding using 384 dimensions and 8 bits per dimension (0.37 KB, 16x smaller).}
    \label{fig:miniatura}
\end{figure}

Applying these techniques introduces an inherent trade-off: reducing the storage size (through lower precision or fewer dimensions) can potentially lead to a loss of information, which might negatively impact the accuracy of the similarity search and, consequently, the overall performance of the RAG system. Figure \ref{fig:miniatura} visually demonstrates this storage challenge and the potential compression achievable through these optimization techniques. Addressing this trade-off effectively requires not only understanding the impact of individual techniques but also their combined effects. Therefore, beyond analyzing these methods, this paper proposes a systematic methodology for selecting the ideal combination of quantization and dimensionality reduction. This methodology aims to identify the specific configuration that maximizes retrieval performance while adhering to specific, predefined memory constraints, thus tackling the core problem of balancing embedding storage efficiency and retrieval quality in RAG systems.

\vspace{-3mm}
\section{Methodology} \label{sec4}

\begin{figure}[htpb]
    \centering
    \includegraphics[width=\textwidth]{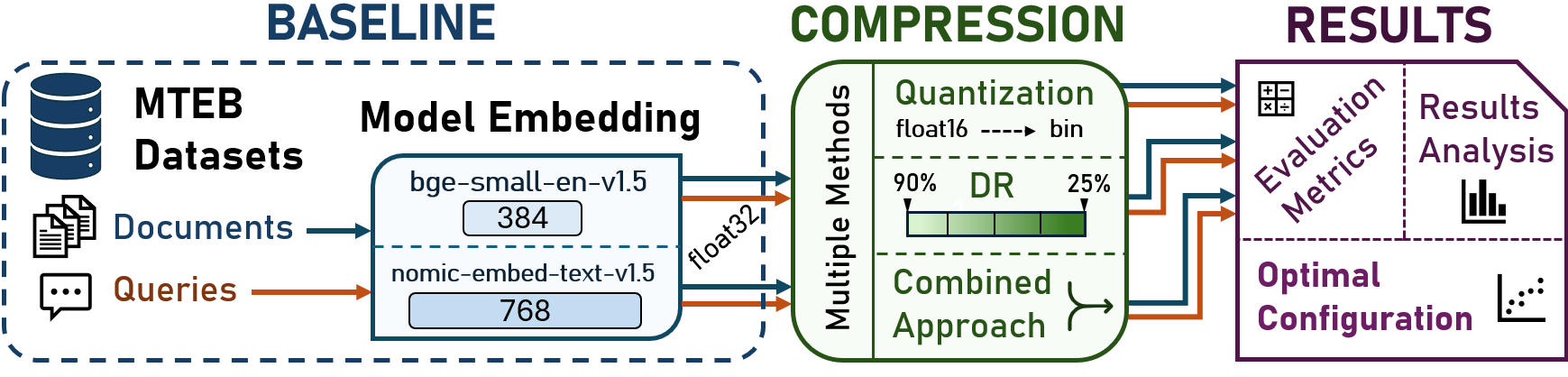} 
    \caption{Experimental pipeline illustrating baseline generation, compression, and evaluation stages.}
    \label{fig:metodologia}
\end{figure}

To systematically evaluate the impact of embedding compression on retrieval performance within RAG systems, we propose a comprehensive experimental framework as illustrated in Figure \ref{fig:metodologia}. This section outlines the baseline construction, compression strategies, and evaluation procedures employed to assess performance degradation resulting from compression.

\subsection{Baseline}

We use the MTEB datasets to ensure standardized evaluation across diverse tasks and domains. Two publicly available models were selected to generate the initial embeddings: \textit{BAAI/bge-small-en-v1.5} (384 dimensions) \cite{bge_embedding} and \textit{Nomic/nomic-embed-text-v1.5} (768 dimensions) \cite{nussbaum2024nomic}, both producing vectors in \textit{float32} precision. The inclusion of models with distinct embedding sizes enables the analysis of how initial dimensionality influences sensitivity to compression.

Thus, baseline embeddings are obtained by processing the MTEB datasets through these models without modification \cite{mteb_arxiv_paper}. Stored and retrieved in their native \textit{float32} format, these embeddings serve as the gold standard against which compressed variants are compared. Retrieval tasks were conducted using the uncompressed document and query embeddings, establishing reference performance metrics.

\subsection{Compression}

Our analysis focuses on two primary strategies: quantization and dimensionality reduction, both evaluated independently and in combination. 

For quantization, we assess several lower-precision formats relative to the \textit{float32} baseline, including \textit{float16}, \textit{bfloat16} \cite{bfloat16deeplearning}, \textit{int8}, and \textit{binary} (1-bit). We also explore novel low-bit floating-point formats, specifically the \textit{float8} variants (\textit{e5m2} and \textit{e4m3} \cite{MLSYS2024_fp8}) and \textit{float4}, implemented via Google \textit{ml\_dtypes} library. These formats vary in their allocation of bits to exponent and mantissa, offering different trade-offs between dynamic range and precision.

Dimensionality reduction is addressed using PCA, Kernel PCA (with cosine, polynomial, and RBF kernels), UMAP, a simple feed-forward autoencoder (two encoder and two decoder layers), and Gaussian Random Projections. For methods requiring model training (PCA, Kernel PCA, and autoencoders), training is conducted on the \textit{MLQuestions} dataset. Reduction levels are set to retain 90\%, 75\%, 50\% and 25\% of the original dimensions.

Combined compression experiments apply a quantization method to vectors resulting from the most promising dimensionality reduction technique identified through preliminary analysis. This enables systematic exploration of trade-offs arising from the concurrent application of both strategies.

\subsection{Evaluation Procedure and Metrics}

Retrieval is performed for each compression setting and model using the corresponding compressed embeddings. Retrieval effectiveness was primarily measured using the Normalized Discounted Cumulative Gain at 10 (nDCG@10), which assesses the relevance and ranking quality of the top 10 retrieved documents. It is formally defined as:

\begin{equation}\label{eq:ndcg10}
nDCG@10 = \frac{DCG@10}{IDCG@10} = \frac{\sum_{i=1}^{10}\frac{rel_i}{\log_2(i+1)}}{IDCG@10}
\end{equation}


where \( \text{IDCG@10} \) is the Ideal Discounted Cumulative Gain at 10, representing the maximum achievable DCG score under a perfect ranking. The logarithmic discount factor penalizes relevant documents appearing lower in the ranking, emphasizing early retrieval of relevant results. Normalization by the ideal score ensures that \( \text{nDCG@10} \) ranges between 0 and 1, with 1 indicating a perfect ranking.

This metric is particularly suitable for evaluating retrieval systems because it emphasizes not only the retrieval of relevant documents but also their placement at the top of the result list, aligning closely with real-world retrieval objectives.

\begin{figure}[htpb]
    \centering
    \includegraphics[width=\textwidth]{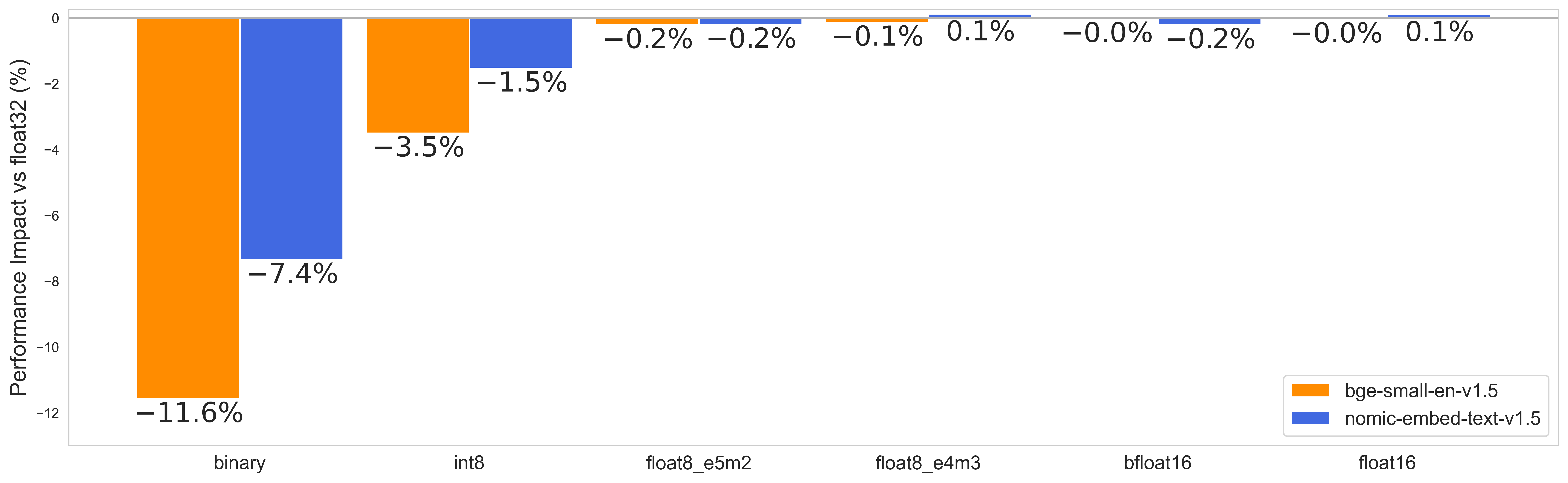} 
    \caption{Impact of quantization on retrieval performance (nDCG@10 relative to float32 baseline) for bge-small-en-v1.5 and nomic-embed-text-v1.5 across different data types.}
    \label{fig:quantization-impact} 
\end{figure}

\vspace{-3mm}
\section{Results}\label{sec5}

This section presents the experimental results evaluating the impact of various quantization and dimensionality reduction techniques on embedding retrieval performance within a RAG context. Performance is measured using \(nDCG@10\) on the MTEB Retrieval benchmark \cite{mteb_arxiv_paper}, reported as the percentage change relative to the \textit{float32} baseline performance for each respective embedding model.

\subsection{Impact of Quantization on Retrieval Performance}
\label{sec:quant_results}

As proposed in Section \ref{sec4}, we first evaluate the effect of applying different quantization techniques, without considering dimensionality reduction. Figure \ref{fig:quantization-impact} reports distinct trade-offs between storage reduction and performance degradation across data types. 

As shown in Figure \ref{fig:quantization-impact}, standard 16-bit floating-point formats (\textit{float16}, \textit{bfloat16}) consistently incurred minimal performance loss compared to the \textit{float32} baseline, while achieving a 2x storage reduction. The novel 8-bit floating-point formats (\textit{float8 e5m2}, \textit{float8 e4m3}) offered an even better compromise, providing a 4x storage reduction with similarly negligible performance degradation. On the other hand, \textit{float4} incurred a very significant performance impact (see Table \ref{tab:combined_results}), and thus will not be considered as an alternative for the remainder of the article.

Scalar quantization formats also exhibit significant performance impacts. The \textit{int8} quantization, despite also offering a 4x storage reduction, results in a considerably higher performance drop (1.5 to 3.5\% depending on the model) compared to \textit{float8} formats. This difference is particularly noteworthy because \textit{int8} quantization requires an additional calibration dataset to determine the minimum and maximum range for each dimension, a step not needed for \textit{float8} formats, which allow direct casting from \textit{float32}. This suggests \textit{float8} is not only higher performing but also simpler to implement. \textit{Binary} quantization, although providing the maximum compression (32x), incurs in a substantial performance loss (ranging from 7\% to over 11\%), often necessitating re-ranking strategies to be viable \cite{Shakir2024}. Comparing the embedding models, the higher-dimensional nomic-embed-text-v1.5 (768 dimensions) generally showed slightly more resilience to quantization, especially for the \textit{binary} format, compared to bge-small-en-v1.5 (384 dimensions), likely due to greater information redundancy.

\subsection{Impact of Dimensionality Reduction Methods}
\label{sec:dim_reduction_results}

After analyzing the effects of quantization, we evaluate the impact of applying various dimensionality reduction techniques to the original embeddings, aiming to identify the most effective methods for preserving retrieval performance while reducing storage. The techniques explored include PCA, Kernel PCA (with cosine, polynomial, and RBF kernels), Uniform Manifold Approximation and Projection (UMAP), Autoencoders, and Gaussian Random Projections. For the Autoencoder, a simple architecture is implemented consisting of two encoder layers and two decoder layers, using the intermediate latent representation as the reduced-dimension embedding.

\begin{figure}[htpb]
    \centering
    \includegraphics[width=\textwidth]{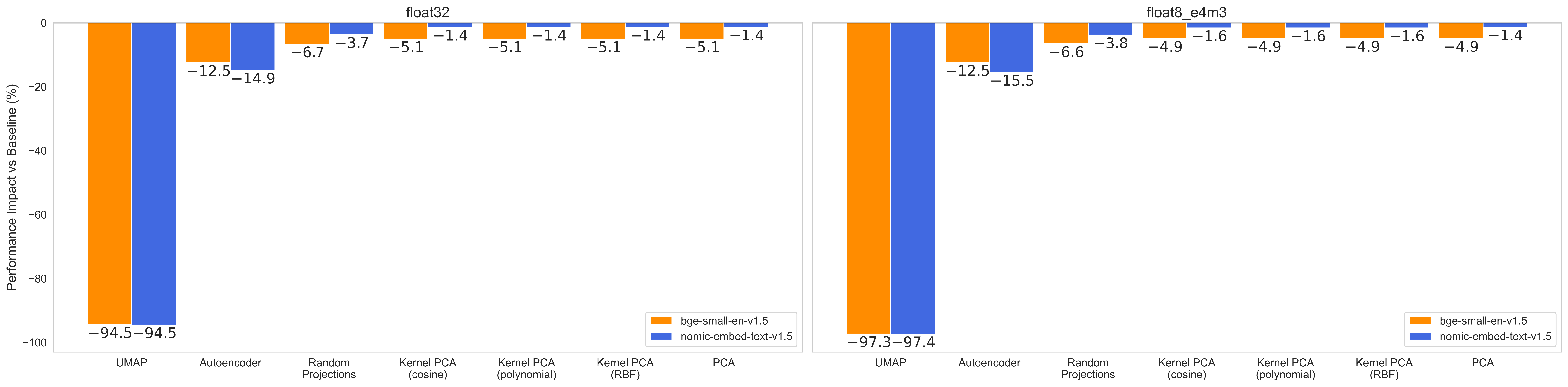}
    \caption{Performance impact of different dimensionality reduction methods (50\% reduction) embeddings for both models. Performance is relative to the respective datatype baseline without reduction.}
    \label{fig:dim_reduction_methods_comparison}
\end{figure}

This comparison was performed for both embedding models and across two representative data types: \textit{float32} and \textit{float8} (\textit{e4m3} variant). This allows us to assess the robustness of each reduction method across different model dimensionalities and data precisions. Figure \ref{fig:dim_reduction_methods_comparison} illustrates the performance impact when reducing the dimensionality by 50\% for \textit{float32} and \textit{float8} respectively. 

The results across both figures and both models consistently show that standard PCA and Kernel PCA (regardless of the kernel) significantly outperform the other methods. These techniques manage to retain a high level of retrieval performance even with substantial dimension reduction. Random Projections yield moderate performance, while Autoencoders and especially UMAP result in drastic performance degradation, indicating they are less suitable for preserving the semantic information required for retrieval.

Crucially, standard PCA achieves performance nearly identical to the best Kernel PCA variants while being computationally much simpler and significantly faster to train and apply. This strong performance holds true for both the higher-dimensional and lower-dimensional models, as well as for both the high-precision \textit{float32} and low-precision \textit{float8} data types. These consistent results strongly suggest that PCA is the most practical and effective dimensionality reduction technique among those tested for this specific task.

\begin{figure}[htpb]
    \centering
    \includegraphics[width=\textwidth]{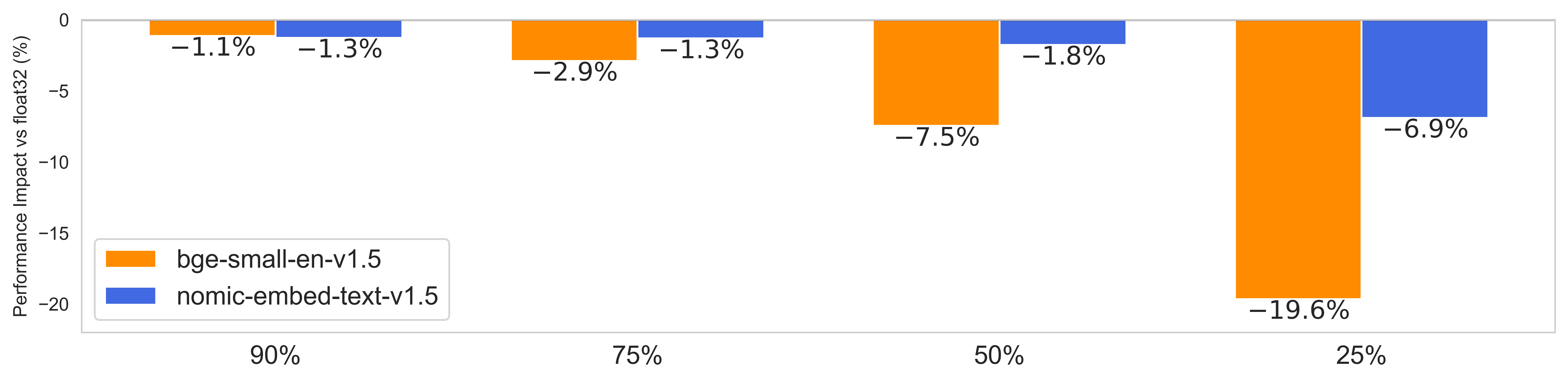} 
    \caption{Performance impact of applying PCA at different reduction levels (90\%, 75\%, 50\%, 25\% dimensions kept) to original \textit{float32} embeddings for both models. Performance is relative to the respective float32 baseline without reduction.}
    \label{fig:pca_impact_float32}
\end{figure}

Having established the suitability of PCA, we further analyze its impact at different reduction levels (retaining 90\%, 75\%, 50\%, and 25\% of the original dimensions) when applied directly to the original \textit{float32} embeddings. Figure \ref{fig:pca_impact_float32} details this analysis. As expected, performance degrades as more dimensions are removed. The higher-dimensional nomic-embed-text-v1.5 model demonstrates greater resilience, with only a -6.9\% drop even at 25\% dimensions retained. Conversely, the bge-small-en-v1.5 model suffers a much larger -19.6\% loss at the same 4x reduction level, highlighting the sensitivity of lower-dimensional embeddings to PCA. Comparing this to quantization results from Section \ref{sec:quant_results}, applying PCA alone to \textit{float32} vectors generally leads to greater performance loss than using quantization formats like \textit{float8} for the same storage reduction factor. This reinforces the idea that PCA, while effective, might be best utilized in combination with quantization rather than as a standalone replacement when starting from high-precision vectors. Based on these findings, PCA is selected as the representative dimensionality reduction technique for analyzing combined effects in the next section.

\subsection{Combined Effects of Quantization and PCA}
\label{sec:combined_results}

Having established the individual effectiveness of quantization and dimensionality reduction, particularly through PCA, we investigate their combined impact in this subsection. This involves applying quantization to vectors whose dimensionality has already been reduced.

\begin{figure}[htpb]
    \centering
    \includegraphics[width=\textwidth]{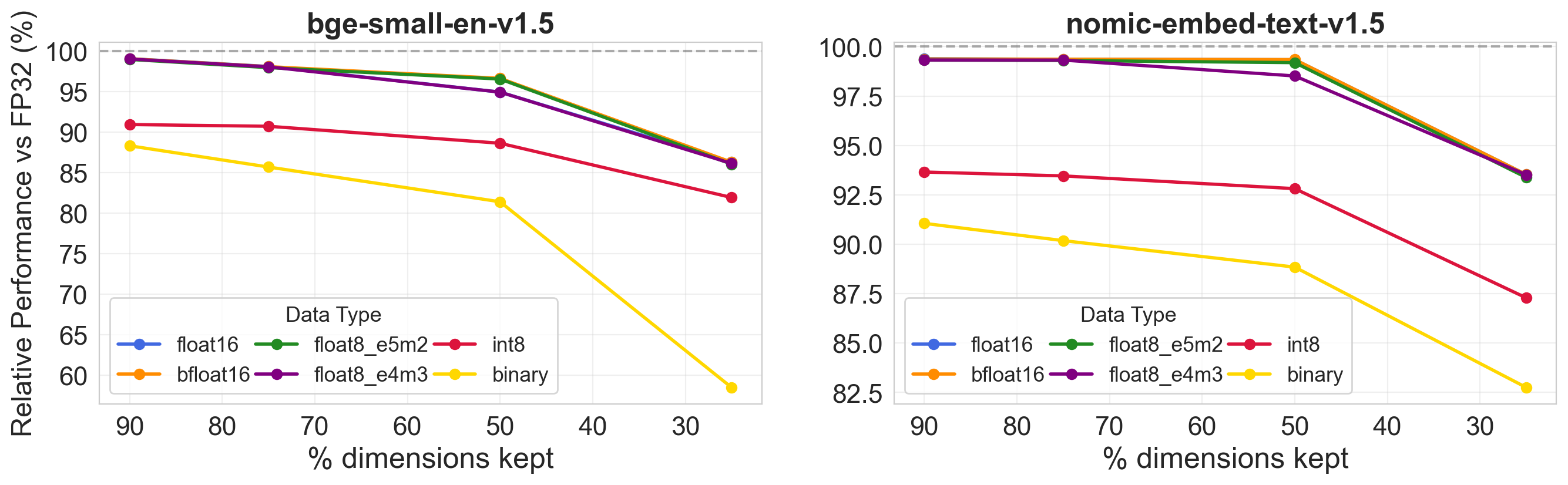} 
    \caption{Relative retrieval performance (nDCG@10 vs float32 baseline) when combining quantization and PCA dimensionality reduction for (a) bge-small-en-v1.5 and (b) nomic-embed-text-v1.5 models. Lines show performance trends for different quantization types as the percentage of dimensions retained via PCA decreases.}
    \label{fig:combined-impact} 
\end{figure}

The interaction between these techniques reveals important synergies and trade-offs. As shown in Figure \ref{fig:combined-impact}, reduced-precision floating-point formats (\textit{float16}, \textit{bfloat16}, \textit{float8}) maintain their robustness even when applied to PCA-reduced vectors. 

Their performance degrades gracefully as dimensionality decreases. \textit{int8} behaves in a similar way but with worse performance, and \textit{binary} exhibits a much steeper decline when combined with PCA.

This combined approach allows for highly efficient configurations. For example, applying \textit{float8} quantization to vectors reduced to 50\% of their original dimensions via PCA (an overall 8x storage reduction) results in a superior performance to using \textit{int8} quantization alone (a 4x reduction). This shows that combining moderate PCA with robust quantization such as \textit{float8} can be a more effective strategy to achieve higher competitive compression ratios than relying solely on aggressive scalar quantization such as \textit{binary}.

Table \ref{tab:combined_results} provides detailed \(nDCG@10\) scores for numerous combinations of embedding model, quantization type, and PCA reduction level across the MTEB Retrieval benchmark datasets, using a weighted average by the number of tokens in the dataset.

\begin{figure}[htpb]
    \centering
    \includegraphics[width=\textwidth]{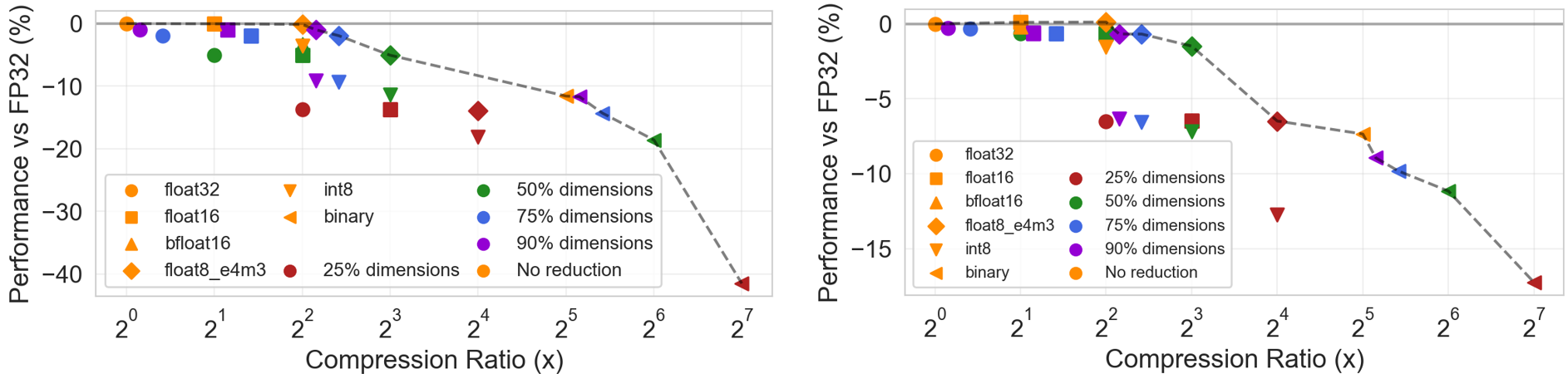} 
    \caption{Performance loss vs. Compression Ratio for \textit{bge-small-en-v1.5} (left) and \textit{nomic-embed-text-v1.5} (right) across all tested quantization and dimensionality reduction combinations. The Y-axis shows the percentage performance drop relative to the \textit{float32} baseline without reduction. The X-axis shows the overall compression ratio achieved (log scale). The dashed line connects the configurations yielding the best observed performance for each compression level within each model.}
    \label{fig:performance_vs_compression} 
\end{figure}

Figure \ref{fig:performance_vs_compression} further synthesizes these trade-offs by plotting the performance loss against the achieved compression ratio (relative to \textit{float32} without compression) for all tested configurations. The dashed line connects the points representing the best achievable performance found in our experiments for each level of compression. This visualization clearly illustrates that for lower compression ratios (up to 4x), direct quantization using robust formats like \textit{float16} or \textit{float8} offers the best performance with minimal loss. However, to achieve higher compression ratios (between 4x and 32x), combining \textit{float8} quantization with varying levels of dimensionality reduction proves to be the optimal strategy, consistently defining the efficiency frontier. At exactly 32x compression, \textit{binary} quantization alone emerges as the best performing option. Beyond 32x, achieving further compression necessitates applying PCA to \textit{binary} quantized vectors, which constitutes the only available path, albeit with a significant performance penalty.


\renewcommand{\arraystretch}{1} 
\setlength{\tabcolsep}{6pt}

\begin{table}[htp]
    \begin{minipage}{.45\textwidth}
      \centering
      {\fontsize{9}{11}\selectfont
        \begin{tabular}{lcccc}
        \toprule
        \multirow{2}{*}{\textbf{dtype}} & \multirow{2}{*}{\textbf{DR}} & \multirow{2}{*}{\textbf{CR}} & \multicolumn{2}{c}{\textbf{nDCG\emph{@}10}} \\ \cmidrule{4-5}
        & & & bge & nomic \\
        \midrule
\multirow{5}{*}{float32} & 100\% & 1.0 & 0.595 & 0.593 \\
 & 90\% & 1.11 & 0.589 & 0.591 \\
 & 75\% & 1.33 & 0.583 & 0.591 \\
 & 50\% & 2.0 & 0.565 & 0.589 \\
 & 25\% & 4.0 & 0.513 & 0.555 \\ \cmidrule{2-5}
\multirow{5}{*}{float16} & 100\% & 2.0 & 0.595 & 0.594 \\
 & 90\% & 2.22 & 0.589 & 0.589 \\
 & 75\% & 2.67 & 0.584 & 0.589 \\
 & 50\% & 4.0 & 0.565 & 0.589 \\
 & 25\% & 8.0 & 0.513 & 0.555 \\ \cmidrule{2-5}
\multirow{5}{*}{bfloat16} & 100\% & 2.0 & 0.595 & 0.592 \\
 & 90\% & 2.22 & 0.589 & 0.589 \\
 & 75\% & 2.67 & 0.583 & 0.589 \\
 & 50\% & 4.0 & 0.575 & 0.589 \\
 & 25\% & 8.0 & 0.513 & 0.555 \\ \cmidrule{2-5}
\multirow{5}{*}{float8\_e4m3} & 100\% & 4.0 & 0.594 & 0.594 \\
 & 90\% & 4.44 & 0.589 & 0.589 \\
 & 75\% & 5.33 & 0.583 & 0.589 \\
 & 50\% & 8.0 & 0.565 & 0.584 \\
 & 25\% & 16.0 & 0.512 & 0.555 \\
        \bottomrule
        \end{tabular}
        }
    \end{minipage}%
    \hspace{0.4cm}
    \begin{minipage}{0.007\textwidth}  
        \centering
    \rule{1pt}{9.6cm}  
    \end{minipage}%
        \begin{minipage}{0.007\textwidth}  
        \centering
    \rule{1pt}{9.6cm}  
    \end{minipage}%
    \hspace{0.35cm}
    \begin{minipage}{.45\textwidth}
      \centering
      {\fontsize{9}{11}\selectfont
        \begin{tabular}{lcccc}
        \toprule
        \multirow{2}{*}{\textbf{dtype}} & \multirow{2}{*}{\textbf{DR}} & \multirow{2}{*}{\textbf{CR}} & \multicolumn{2}{c}{\textbf{nDCG\emph{@}10}}\\ \cmidrule{4-5}
        & & & bge & nomic \\
        \midrule
\multirow{5}{*}{float8\_e5m2} & 100\% & 4.0 & 0.593 & 0.592 \\
 & 90\% & 4.44 & 0.589 & 0.589 \\
 & 75\% & 5.33 & 0.583 & 0.589 \\
 & 50\% & 8.0 & 0.574 & 0.588 \\
 & 25\% & 16.0 & 0.512 & 0.554 \\ \cmidrule{2-5}
\multirow{5}{*}{float4\_e2m1} & 100\% & 8.0 & 0.0 & 0.406 \\
 & 90\% & 8.89 & 0.002 & 0.001 \\
 & 75\% & 10.67 & 0.002 & 0.001 \\
 & 50\% & 16.0 & 0.002 & 0.001 \\
 & 25\% & 32.0 & 0.002 & 0.001 \\ \cmidrule{2-5}
\multirow{5}{*}{int8} & 100\% & 4.0 & 0.574 & 0.584 \\
 & 90\% & 4.44 & 0.541 & 0.555 \\
 & 75\% & 5.33 & 0.54 & 0.554 \\
 & 50\% & 8.0 & 0.527 & 0.55 \\
 & 25\% & 16.0 & 0.487 & 0.518 \\ \cmidrule{2-5}
\multirow{5}{*}{binary} & 100\% & 32.0 & 0.526 & 0.549 \\
 & 90\% & 35.56 & 0.525 & 0.54 \\
 & 75\% & 42.67 & 0.51 & 0.535 \\
 & 50\% & 64.0 & 0.484 & 0.527 \\
 & 25\% & 128.0 & 0.348 & 0.491 \\
        \bottomrule
        \end{tabular}
        }
    \end{minipage}%
    \vspace{0.25cm}
\caption{Detailed performance (nDCG@10) for combined Quantization and PCA levels. Weighted average across MTEB Retrieval datasets, where DR represents the dimensionality reduction and CR the compression rate applied.}
\label{tab:combined_results}
\end{table}

\subsection{Methodology for Optimal Configuration Selection}
\label{sec:methodology}

The extensive results of this manuscript highlight the complex interplay between embedding model choice, quantization type, and dimensionality reduction level. Selecting the optimal configuration requires balancing retrieval performance against storage constraints specific to the target application. To facilitate this decision-making process, we propose a methodology based on visualizing the performance-storage trade-off space.

This involves plotting the achieved retrieval performance (e.g., average \(nDCG@10\) across MTEB, or a dataset-specific score) 
against the required storage size (e.g., in MB for a representative number of embeddings, like 100k or 1M). 
Each point on this plot represents a unique configuration (specific embedding model, quantization type, and PCA reduction level). The storage size for \(N\) embeddings is calculated as: \( \text{Storage (bytes)} = N \times (\text{Original Dimensions} \times \text{PCA Ratio}) \times \text{Bytes per Dimension} \). 

\begin{figure}[t!]
    \centering
    \includegraphics[width=\textwidth]{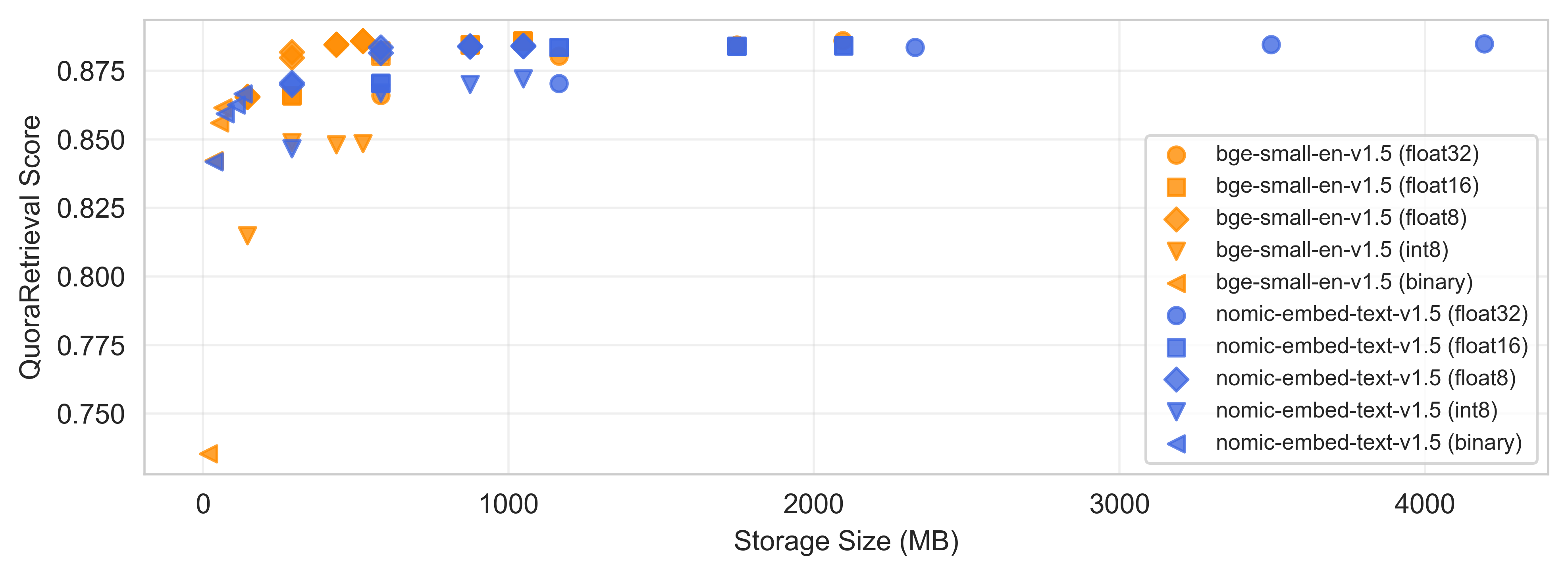} 
    \caption{Storage Size vs. QuoraRetrieval Performance for various configurations. Vertical lines indicate typical memory zones for different device classes. Allows selection of the best performing configuration within a given memory budget.}
    \label{fig:tradeoff_plot_quora}
\end{figure}

\begin{figure}[b!]
    \centering
    \includegraphics[width=\textwidth]{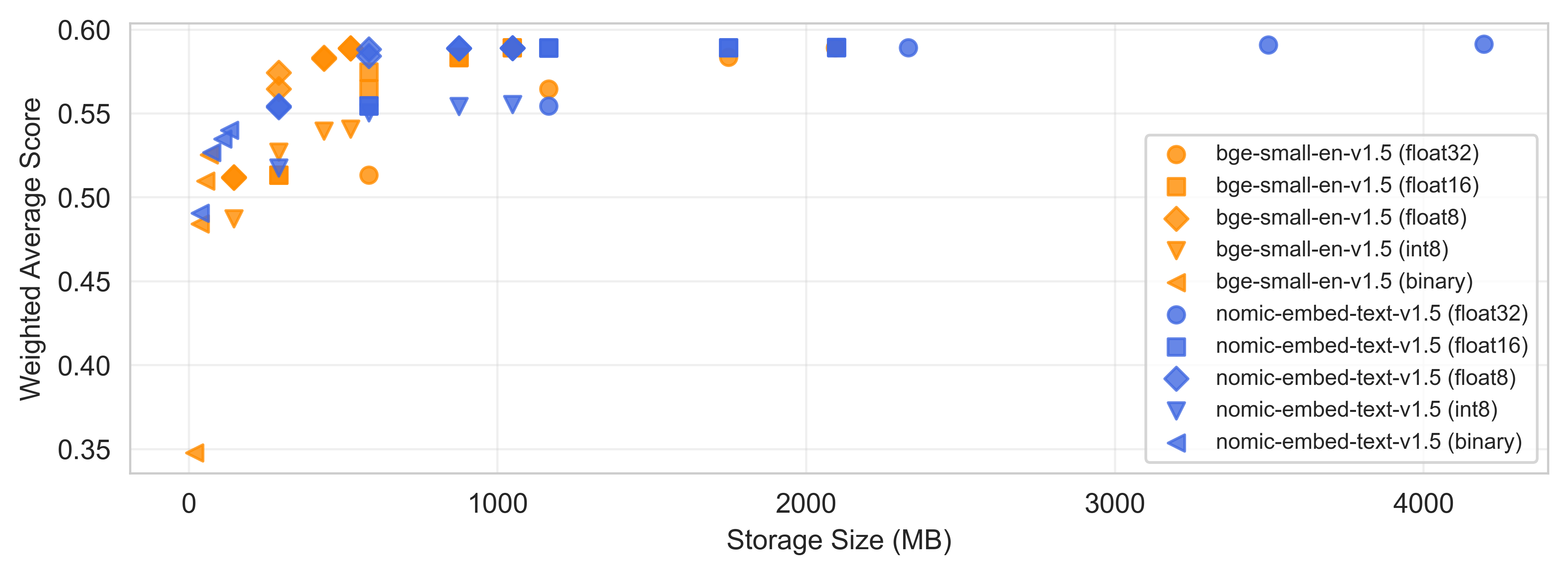} 
    \caption{Storage Size vs. MTEB Weighted Average Performance for a hypothetical 125k embedding dataset. Illustrates how optimal configurations shift based on dataset size and available memory.}
    \label{fig:tradeoff_plot_125k}
\end{figure}

By plotting the achieved retrieval performance 
versus the required storage size, 
practitioners can visualize the performance-storage trade-off space for various configurations. Each point represents a unique combination of embedding model, quantization type, and PCA reduction level. Figures \ref{fig:tradeoff_plot_quora} and \ref{fig:tradeoff_plot_125k} illustrate this visualization approach for the specific QuoraRetrieval dataset and a hypothetical 125k embedding dataset using MTEB average scores, respectively. 

The selection process is straightforward: identify the target memory constraint line, 
among all configuration points lying to the left of this line (i.e., meeting the memory budget), select the one with the highest performance score (highest Y-value). This point represents the Pareto-optimal configuration for that specific memory limit.

This methodology can be adaptable. While Figures \ref{fig:tradeoff_plot_quora} and \ref{fig:tradeoff_plot_125k} use MTEB averages or a specific dataset score, the same technique can be applied using performance metrics from any relevant benchmark or internal evaluation dataset, allowing for highly tailored optimization based on the specific data and performance requirements of the target RAG application.

\vspace{-3mm}
\section{Conclusions}\label{sec6}

This work investigates the impact of various quantization and dimensionality reduction techniques on the storage efficiency and retrieval performance of embeddings within RAG systems. Our comprehensive evaluation across different embedding models, data types, reduction methods, and compression levels yields several key conclusions for optimizing embedding storage:

\begin{itemize}
    \item Reduced-precision floating-point formats, particularly \textit{float8} variants (\textit{e4m3} and \textit{e5m2}), offer an excellent trade-off between precision and compression, achieving 4x storage reduction with minimal performance degradation. 
    They significantly outperform standard \textit{int8} quantization at the same compression level 
    and possess the added advantage of not requiring a separate data calibration step. 
    This questions the common industry practice of defaulting to \textit{int8} for 4x compression.
    
    \item 
    Standard PCA and Kernel PCA consistently demonstrate the best performance in preserving retrieval quality. Given that standard PCA offers comparable results to Kernel PCA with significantly lower computational overhead for training, it emerges as the most practical choice for dimensionality reduction in RAG systems.
    
    \item Combining moderate PCA with robust quantization formats like \textit{float8} enables higher compression ratios (e.g., 8x or more) while maintaining better retrieval performance than resorting to either aggressive dimensionality reduction on \textit{float32} vectors or quantization 
    alone. 
    Additionally, floating-point formats consistently show greater resilience to combined compression compared to scalar formats (\textit{int8}, \textit{binary}).
    
    \item Higher-dimensional embedding models consistently exhibited greater resilience to both quantization and dimensionality reduction compared to their lower-dimensional counterparts, degrading more gracefully under compression. This suggests that the inherent information redundancy in higher-dimensional spaces provides a buffer against information loss during compression.
    
    \item The proposed visualization methodology, plotting retrieval performance against storage size for various configurations, provides a clear and adaptable framework for practitioners. It allows for the identification of Pareto-optimal configurations that maximize performance within specific memory budgets, tailored to the requirements of diverse deployment scenarios.
\end{itemize}

These findings provide actionable insights for optimizing embedding storage in RAG systems. By leveraging efficient formats like \textit{float8} and strategically combining them with PCA, practitioners can significantly reduce memory footprint and associated costs, enabling more scalable and efficient deployments without substantial performance sacrifices.

\vspace{-3mm}
\section*{Acknowledgments}
This work has been funded by the project of the Spanish Ministry of Science and Innovation grant number PID2023-153047OBI00, by the Consejería de Educación de la Junta de Castilla y León and by the Universidad de León.

\bibliographystyle{plain}

\vspace{-0.15in}

\bibliography{bibliography}

\begin{thebibliography}{10}

\bibitem{ieee754}
IEEE.
\newblock 754-2019 - ieee standard for floating-point arithmetic, July 2019.

\bibitem{kusupati2024matryoshkarepresentationlearning}
Aditya Kusupati, Gantavya Bhatt, Aniket Rege, Matthew Wallingford, Aditya Sinha, Vivek Ramanujan, William Howard-Snyder, Kaifeng Chen, Sham Kakade, Prateek Jain, and Ali Farhadi.
\newblock Matryoshka representation learning.
\newblock {\em Advances in Neural Information Processing Systems}, 35:30233--30249, 12 2022.

\bibitem{ragfacebook}
Patrick Lewis, Ethan Perez, Aleksandra Piktus, Fabio Petroni, Vladimir Karpukhin, Naman Goyal, Heinrich Küttler, Mike Lewis, Wen tau Yih, Tim Rocktäschel, Sebastian Riedel, and Douwe Kiela.
\newblock Retrieval-augmented generation for knowledge-intensive nlp tasks.
\newblock {\em Advances in Neural Information Processing Systems}, 33:9459--9474, 2020.

\bibitem{bfloat16deeplearning}
Saras~Mani Mishra, Ankita Tiwari, et~al.
\newblock Comparison of floating-point representations for the efficient implementation of machine learning algorithms.
\newblock In {\em 2022 32nd International Conference Radioelektronika (RADIOELEKTRONIKA)}, pages 1--6, 2022.

\bibitem{mteb_arxiv_paper}
Niklas Muennighoff, Nouamane Tazi, Loïc Magne, and Nils Reimers.
\newblock Mteb: Massive text embedding benchmark.
\newblock 10 2022.

\bibitem{nussbaum2024nomic}
Zach Nussbaum, John~X. Morris, Brandon Duderstadt, and Andriy Mulyar.
\newblock Nomic embed: Training a reproducible long context text embedder, 2024.

\bibitem{llm_hallucinations}
Gabrijela Perković, Antun Drobnjak, and Ivica Botički.
\newblock Hallucinations in llms: Understanding and addressing challenges.
\newblock {\em 2024 47th ICT and Electronics Convention, MIPRO 2024 - Proceedings}, pages 2084--2088, 2024.

\bibitem{reimers2019sentenceembeddings}
Nils Reimers and Iryna Gurevych.
\newblock Sentence-bert: Sentence embeddings using siamese bert-networks.
\newblock {\em EMNLP-IJCNLP 2019 - 2019 Conference on Empirical Methods in Natural Language Processing and 9th International Joint Conference on Natural Language Processing, Proceedings of the Conference}, pages 3982--3992, 2019.

\bibitem{Shakir2024}
Aamir Shakir, Tom Aarsen, and Sean Lee.
\newblock Binary and scalar embedding quantization for significantly faster and cheaper retrieval, 3 2024.

\bibitem{MLSYS2024_fp8}
Haihao Shen, Naveen Mellempudi, et~al.
\newblock Efficient post-training quantization with fp8 formats.
\newblock In P.~Gibbons, G.~Pekhimenko, and C.~De Sa, editors, {\em Proceedings of Machine Learning and Systems}, volume~6, pages 483--498, 2024.

\bibitem{ft_vs_rag}
Heydar Soudani, Evangelos Kanoulas, and Faegheh Hasibi.
\newblock Fine tuning vs. retrieval augmented generation for less popular knowledge.
\newblock In {\em Proceedings of the 2024 Annual International ACM SIGIR Conference on Research and Development in Information Retrieval in the Asia Pacific Region}, SIGIR-AP 2024, page 12–22, New York, NY, USA, 2024. Association for Computing Machinery.

\bibitem{model_compression_overview}
Ching~Hao Wang, Kang~Yang Huang, Yi~Yao, Jun~Cheng Chen, Hong~Han Shuai, and Wen~Huang Cheng.
\newblock Lightweight deep learning: An overview.
\newblock {\em IEEE Consumer Electronics Magazine}, 13:51--64, 7 2024.

\bibitem{embeddings_apple_pca}
Yu~Wang.
\newblock Single training dimension selection for word embedding with {PCA}.
\newblock {\em CoRR}, abs/1909.01761, 2019.

\bibitem{ragsurvey}
Shangyu Wu, Ying Xiong, Mbzuai~Yufei Cui, Haolun Wu, Can Chen, Ye~Yuan, Lianming Huang, Xue Liu MBZUAI Tei-Wei Kuo, Nan Guan, Chun Jason~Xue MBZUAI, Yufei Cui, Xue Liu, Tei-Wei Kuo, and Chun~Jason Xue.
\newblock Retrieval-augmented generation for natural language processing: A survey.
\newblock {\em ArXiv}, 7 2024.

\bibitem{bge_embedding}
Shitao Xiao, Zheng Liu, Peitian Zhang, and Niklas Muennighoff.
\newblock C-pack: Packaged resources to advance general chinese embedding, 2023.

\end{thebibliography}

\end{document}